\documentclass{aa}
\usepackage{natbib}
\usepackage{xspace}
\usepackage{amssymb}
\usepackage{amsmath}
\usepackage{graphicx}
\def\cdrev#1{{{#1}}}
\def\revised#1{{{#1}}}

\def\aap{A\& A}

\def\apj{ApJ}

\def\apjl{ApJL}
\def\apjs{ApJS}

\def\fullstop{\,.}

\def\um{\ensuremath{\mu\mathrm{m}}\xspace}

\begin{document}
\title{Size-sorting dust grains in the surface layers of protoplanetary disks}
\titlerunning{Size-sorting dust grains in protoplanetary disks}
\authorrunning{Dullemond \& Dominik} 
\author{C.P.~Dullemond \& C.~Dominik}
\institute{Max Planck Institut f\"ur Astronomie, K\"onigstuhl 17, 69117
Heidelberg, Germany \and Sterrenkundig Instituut `Anton Pannekoek',
Kruislaan 403, NL-1098 SJ Amsterdam, The Netherlands; e--mail:
dominik@science.uva.nl}
\date{DRAFT, \today}

\abstract{
{\em Context:} The shape of dust emission features measured from
protoplanetary disks contains information about the typical size of
the dust particles residing in these disks. A flattened 10 $\mu$m
silicate feature is often interpreted as proof that grain growth
has taken place, while a pointy feature is taken as evidence for the
pristine nature of the dust.
\\
{\em Aims:} We wish to investigate what the effect of dust sedimentation
is on the observed 10 $\mu$m feature and how this may affect the
interpretation of the observations.
\\
{\em Methods:} Using a combination of modeling tools, we simulated the
sedimentation of a dust grain size distribution in an axisymmetric 2-D model
of a turbulent protoplanetary disk, and we used a radiative transfer program
to compute the resulting spectra.
\\
{\em Results:} We find that the sedimentation can turn a flat feature into a
pointy one, but only to a limited degree and for a very limited set of 
particle size distributions. If the distribution is too strongly dominated
by small grains, then the feature is pointy even before sedimentation. If the
distribution is too strongly dominated by big grains, the sedimentation will
not be enough to cause the feature to be pointy. Only if we have a bimodal
size distribution, i.e.~a very small grain population and a bigger grain
population, do we find that the transformation from a flat to a pointy feature
upon dust sedimentation is strong. However, our model shows that, if
sedimentation is the sole reason for the variety of silicate feature
strengths observed in protoplanetary disks, then we would expect to find
a correlation such that disks with weak mid- to far-infrared excess have
a stronger 10 $\mu$m silicate feature than disks with a strong mid- to
far-infrared excess. If this is contrary to what is observed, then this would
indicate that sedimentation cannot be the main reason for the variety of
10 $\mu$m silicate features observed in protoplanetary disks.
}

\maketitle


\section{Introduction}
The protoplanetary disks surrounding Brown Dwarfs, T Tauri stars, and Herbig
Ae/Be stars have displayed a rich variety of dust emission features in their
infrared spectra. These features have been studied in great detail with the
Infrared Space Observatory (ISO; e.g.~Meeus et
al.~\citeyear{meeuswatersbouw:2001}; Bouwman \citeyear{bouwmanmeeus:2001}),
with ground-based instruments (van Boekel et
al.~\citeyear{vanboekelwaters:2003}; Honda et
al.~\citeyear{hondakataza:2003}) and with the Spitzer Space Telescope
(e.g.~Apai et al.~\citeyear{apaiscience:2005}; Kessler-Silacci et
al.~\citeyear{kessleraugereau:2006}; Furlan et
al.~\citeyear{furlanhartmann:2006}). While the wavelength and width of these
dust emission features reveals the composition of the dust grains and
whether they have undergone thermal processing or not, the shape and
strength of the features betrays the dominant size of the emitting
grains. Infrared spectroscopy is therefore a powerful tool for studying the
onset of grain growth, which is believed to be the very initial step toward
planet formation.

A drawback of this kind of analysis is that it only probes the very
tenuous `superheated' surface layers of the disk. This layer, which is
produced by the irradiation of the disk by the central star (Calvet et
al.~\citeyear{calvetpatino:1991}; Chiang \& Goldreich
\citeyear{chianggold:1997}), is very optically thin and has a
much higher temperature than the disk
interior. For these reasons the dust in this layer typically produce
strong emission features. Although this surface layer contains only a
minuscule part of the mass of the disk, it is responsible for about
half of the infrared luminosity of the disk (Chiang \& Goldreich
\citeyear{chianggold:1997}) and entirely dominates the
  generation of dust emission features.  This raises the concern about how
representative this layer is for the entire disk, \cdrev{and whether
  changes to dust properties observed in this layer are in any way
  related to changes deeper in the disk, where most of the mass
  resides and important processes such as planet formation take
  place.} Of particular concern is that dust tends to
sediment toward the midplane until an equilibrium between
sedimentation and vertical turbulent mixing is established (Dubrulle
et al.~\citeyear{dubmorster:1995}).  Large grains have a lower
surface-to-mass ratio and are therefore less prone to friction with
the gas than small grains. They therefore settle more deeply into the
disk interior before an equilibrium between sedimentation and
turbulent mixing is established. Small grains, however, stay afloat
much higher up in the photosphere of the disk, and since they dominate
the opacity anyway, they maintain the photospheric disk height at a
higher level. The bigger grains therefore settle out of the
photosphere, into the deep optically thick regions of the disk. They
are therefore filtered out of sight.

The question therefore arises: Could it be that the dust in this surface
layer is very {\em atypical} compared to the bulk of the disk deeper in the
optically thick parts of the disk?  If this is the case, what will we learn
from studying the infrared spectra from these disks? This paper is not the
first to raise this question. It has also been noted by e.g.\ van Boekel et
al.~(\citeyear{vanboekelwaters:2003}), and Sicilia-Aguilar et
al.~(\citeyear{sicilia:2007}) have invoked the potential effect of big
grains sinking out of the surface layer in their analysis of Spitzer spectra
of a sample of pre-main-sequence stars. Kessler-Silacci et
al.~(\citeyear{kessleraugereau:2006}) speculate that, by looking at
different wavelength ranges, one looks deeper into the disk toward regions
of larger grains. However, no quantitive study of this problem has been
presented yet. A related problem has, however, been studied recently by
Dullemond et al.~(\citeyear{dullemond:2007}). That paper showed that dust
sedimentation can enhance the infrared features from polycyclic aromatic
hydrocarbons (PAHs) in the disk, because the sedimentation causes the
thermal dust grains to sink below the photosphere held up by the PAHs.

In the present paper we wish to study the effect of sedimentation on the
shape of the 10 $\mu$m feature of silicate dust. This solid state infrared
emission feature, seen in the majority of protoplanetary disks, has 
often been used as a probe of grain size (e.g.~van Boekel et
al.~\citeyear{vboekelmin:2005}; Schegerer et
al.~\citeyear{schegerer:2006}). A strong and pointy 10 $\mu$m feature is
typical of grains smaller than about 1 $\mu$m in radius, while a weak and
flattened feature indicates the presence of grains between 2 $\mu$m and
4$\mu$m, clearly larger than the average grain size in the interstellar
medium. Van Boekel et al.~(\citeyear{vanboekelwaters:2003}) showed that not
only are such variations in feature shape and strength observed in
protoplanetary disks, but indeed also the correlation between shape and
strength expected from the laboratory measurements is observed. We are aware
that the interpretation of the feature shape in terms of grain size is still
under debate, as porosity and fractal structure of the aggregates may
influence the absorption/emission cross section (Voshchinnikov et
al.~\citeyear{voshch:2006}; Min et al.~\citeyear{minhove:2005}). Moreover,
Juhasz et al.~(A\&A submitted) has shown that high signal-to-noise is needed
to interpret these features in the first place. But the fact that
observations appear to confirm the predicted trend gives some hope that the
method of interpretation of these features is reasonable.

\cdrev{Many attempts have been made to link the shape and strength of the 10
  $\mu$m feature to the evolution of the disk as a whole.  But not even the
  direction of this trend is clear.  Does the a flat, broad feature indicate
  a more developed disk by showing that larger grains have been produced in
  the disk?  This is the most common interpretation in the literature, and
  it keeps the assumption alive that growth from submicron grains to
  micron-sized grains is a process that happens slowly in the course of a
  million years.  However, such a trivial, direct connection seems rather
  unlikely, given the speed at which coagulation can proceed and produce
  large dust grains (Dullemond \& Dominik \citeyear{duldom:2005}).  In view
  of dust settling, the exact opposite relation with the developmental state
  of the disk is possible as well.  Assuming a steady-state size
  distribution and an equilibrium state between settling and turbulent
  mixing, the amount of large grains still present in the surface layer can
  be a simple effect of the strength of turbulent vertical mixing in the
  disk.  As turbulence and gas densities die down toward the end of the disk
  evolution, large grains would tend to disappear from the observer's view
  and result in pointy features for {\em evolved} disks.  For interpreting this
  type of spectroscopy and linking it to the evolutionary state of a disk,
  it is vital to understand if simple settling can indeed filter out large
  grains efficiently.  To establish or disprove this possibility is the core
  purpose of this paper.}

To study the effect of sedimentation on the 10 $\mu$m silicate feature shape,
we used the same setup as in the above mentioned paper on PAHs (Dullemond et
al.~\citeyear{dullemond:2007}). We set up a disk structure in the same way
as described in Dullemond \& Dominik (\citeyear{duldomsett:2004}). We then
assumed some grain size distribution throughout the disk and solve the
time-dependent 1-D vertical dust sedimentation-mixing equations at each
radial grid point for each grain size. Finally we used the Monte Carlo
continuum radiative transfer code {\tt RADMC} to produce the infrared
spectra, focusing on the region between 6 $\mu$m and 15 $\mu$m. We then
analyzed these results and figured out if dust sedimentation can modify the
feature shape. In particular we investigated if sedimentation can turn a
flat feature indicative of large grains into a pointy feature indicative of
small grains.

The structure of this paper is as follows. In Section \ref{sec-opacities} we
discuss the opacities we use and how we construct grain size distributions.
We then discuss the model in Section \ref{sec-model} and the results in
Section \ref{sec-results}. Finally we conclude in Section
\ref{sec-conclusion}.

\section{Opacities and grain size distributions}
\label{sec-opacities}
The results of our analysis may depend on the precise details of the
opacities and grain size distributions used. Therefore we discuss them here
in some detail. We used $n$ and $k$ optical constants from the Jena database
for Mg Fe SiO$_4$. We produced the absorption cross-section using a Mie
calculation and a specific weight of $\xi=$3.6 g/cm$^3$. We did not include
scattering in the radiative transfer calculations. With these Mie
calculations we obtained an opacity for a grain of a specific radius $a$. 

\begin{figure}
\centerline{\includegraphics[width=0.45\textwidth]{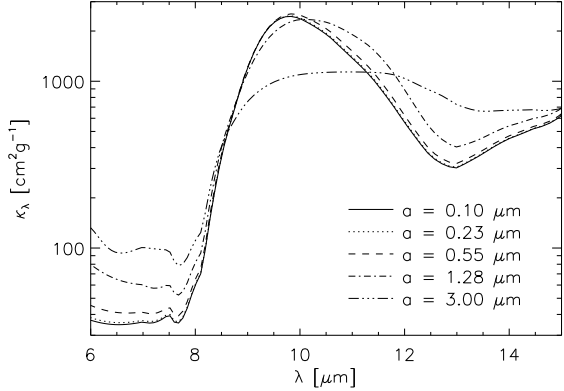}}
\caption{The absorption cross-section in cm$^2$/gram-of-dust in the
N-band range for the dust we use in our model for various spherical
grain radii. For grain radii smaller than 0.1 $\mu$m, the opacity does
not change much from the opacity of 0.1 $\mu$m grains in this wavelength
domain. 
\label{fig-opacities}}
\end{figure}

To model a grain size distribution we must choose a discrete set of grain radii
between the minimum radius $a_{\mathrm{min}}$ and the maximum radius
$a_{\mathrm{max}}$ of the distribution.  Throughout this paper we
choose $a_{\mathrm{min}}=0.001\;\mu$m and $a_{\mathrm{max}}=3\;\mu$m.  In
our model we always choose the sampling points to be logarithmically divided
between these bounds, such that we have linear spacing in log$(a)$.
Figure \ref{fig-opacities} shows the opacities for a limited set of grain
sizes, just to show the shape change occurring when one goes from small
($a\lesssim 0.3\mu$m) to big $a\gtrsim 2\mu$m grains. \revised{It should be
  noted that flat-topped (boxy) features can also be produced by certain
  mixtures of small crystalline silicate grains (e.g.\ Honda et al.\
  \citeyear{hondakataza:2003}). The flatness of the feature is then due to
  the summing of a number of narrow crystalline peaks lined up so that,
  taken together, they look like a broad flat-topped feature of an amorphous
  large grain.  High-resolution and -sensitivity measurements, however,
  would be able to distinguish these two scenarios by identifying the slight
  bumpy shape of the flat-topped feature in the crystalline silicate
  scenario, where the bumps are at the known locations of the peaks of 
  enstatite and forsterite.}

To obtain the total cross section per gram of dust we need to speficy the
grain size distribution function $f(a)$. We define $f(a)$ such that
\begin{equation}
\rho_{\mathrm{dust}}=\int_{a_{\mathrm{min}}}^{a_{\mathrm{max}}}
m(a) f(a) da
\end{equation}
is the total dust density in units of gram/cm$^3$. Here $m(a) = (4\pi/3)\xi
a^3$ with $\xi=$3.6 g/cm$^3$. Throughout this paper we take a powerlaw
distribution: $f(a)\propto a^p$. The case $p=-7/2$ represents the MRN
distribution (Mathis, Rumpl \& Nordsieck \citeyear{mrn:1977}).  In terms of
a size distribution in particle mass $\tilde f(m)\propto m^q$ such that
$\tilde f(m)dm=f(a)da$, one gets $q=(p-2)/3$, so that for the MRN
distribution we have $q=-11/6$. However, from now on we shall use the $f(a)$
formalism. If we sample this distribution function with $N$ discrete
$a$-values, then we can associate a dust density to each of these `bins':
\begin{equation}
\rho_{\mathrm{dust}}^{(i)}=m(a_i)a_if(a_i)\Delta\log(a)_i
\end{equation}
where $\Delta\log(a)_i$ is the width of each bin in $\log(a)$ which we
assume to be the same for all bins, except for the first and the last,
which are only half as wide to ensure that the distribution goes
exactly from $\log(a_{\mathrm{min}})$ to $\log(a_{\mathrm{max}})$. The total
radiative cross section $\alpha_\nu$ in units of $1/$cm is then
\begin{equation}
\alpha_\nu = \sum_{i=1,N} \rho_{\mathrm{dust}}^{(i)} \kappa_\nu^{(i)}\fullstop
\end{equation}
The total emissivity $j_\nu$ in units of
erg$\,$s$^{-1}$$\,$cm$^{-3}$$\,$Hz$^{-1}$$\,$ster$^{-1}$ is then
\begin{equation}
j_\nu = \sum_{i=1,N} \rho_{\mathrm{dust}}^{(i)} \kappa_\nu^{(i)}
B_\nu(T_i)
\end{equation}
where $T_i$ is the temperature of grains with size $a_i$. We allow the
grains of different sizes to obtain different temperatures, because we
assume that they are not mutually thermally coupled, nor do they have strong
thermal coupling with the gas.

Because each grain size can have a different temperature, the $\alpha_\nu$
is not really a correct representation of the expected output spectrum, because the
Planck functions with which the individual $\rho_{\mathrm{dust}}^{(i)}
\kappa_\nu^{(i)}$ are multiplied may be different for each $i$. The full
radiative transfer calculations described later in this paper will take care
of this.

\section{Model}\label{sec-model}
We start our model with making a vertical structure calculation of a disk
around a Herbig Ae star with $T_{*}=10000$K, $R_{*}=2.5R_{\odot}$, $M_{*}
=2.5 M_{\odot}$. The disk has an inner radius of $0.7$ AU, an outer radius
of 100 AU, and a surface density distribution of
$\Sigma(R)=\Sigma_0(R/\mathrm{AU})^{-1}$ with $\Sigma_0=163$g/cm$^2$.  We
then use the procedures described in Dullemond \& Dominik
(\citeyear{duldomdisk:2004}) to create a self-consistent disk structure
that is in vertical hydrostatic equilibrium. This procedure involves
solving the radiative transfer equation in 2-D axisymmetry using a
multi-dimensional radiative transfer program, and iterating this with the
vertical hydrostatic equilibrium equation. This procedure then gives
$\rho_{\mathrm{gas}}(r,z)$, i.e.\ the gas density as a function of radius
and vertical height above the midplane. In this procedure we assume that the
gas and the dust are well-mixed and thermally coupled (i.e.\
$T_{\mathrm{gas}}=T_{\mathrm{dust}}$) and we assume that the dust consists
of small $0.1\mu$m size silicate grains. 

Once we have this gas density distribution we insert the real dust density
distribution that we wish to model, again with the same dust-to-gas ratio of
1:100. We then use the time-dependent dust settling code described in detail
in Dullemond \& Dominik (\citeyear{duldomsett:2004}). This gives the density
$\rho_i(r,z,t)$ of each dust component $i$. For a bimodal distribution we
have only $i=1,2$, while for a continuous distribution we have $i=1\cdots
20$, as we model the distribution with 20 bins. We take the time $t$ to
be 1 Myr, so that we are sure that the dust has settled in a settling-mixing
equilibrium.

\section{Results}\label{sec-results}
\begin{figure*}
\centerline{\includegraphics[width=0.8\textwidth]{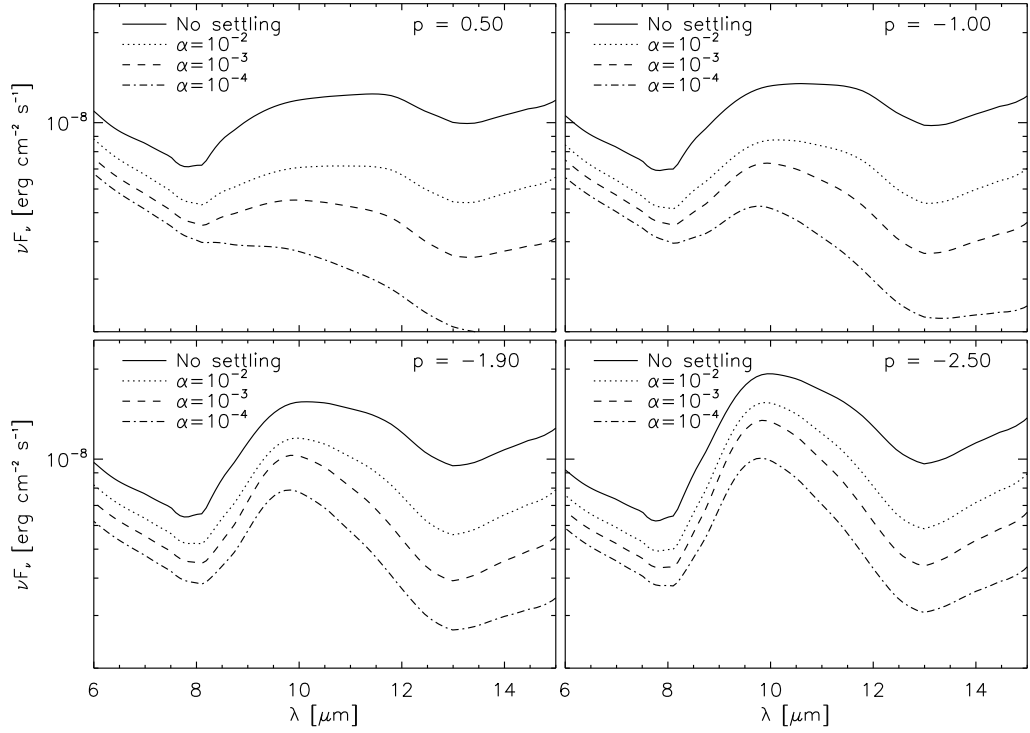}}
\caption{Model results for the continuous size distribution with smallest
  size 0.001 $\mu$m and the biggest size 3 $\mu$m. The four different panels
  are for four different values of the powerlaw index $p$ for the grain size.
  The four different lines in each
  panel are for no settling (top) and for settling-mixing equilibrium with
  three different levels of turbulence. 
\label{fig-result-spec-distr}}
\end{figure*}

\begin{figure*}
\includegraphics[width=\textwidth]{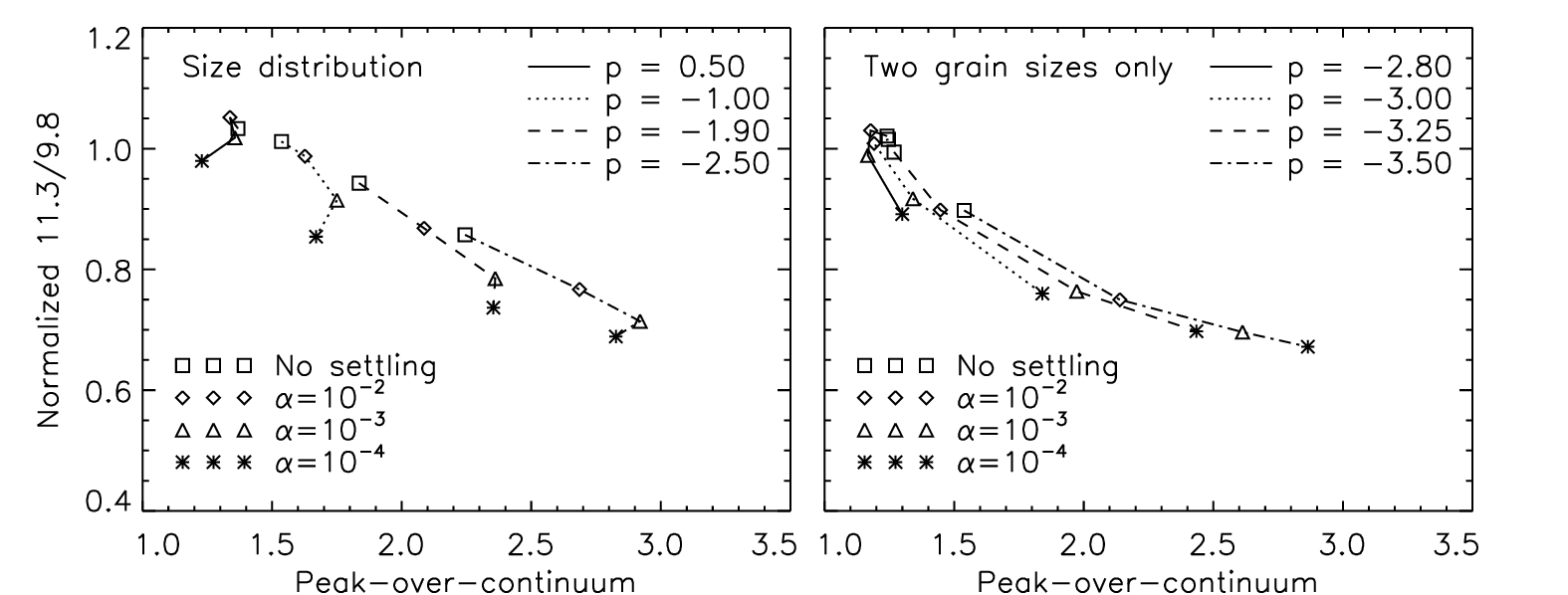}
\caption{The feature-strength versus shape diagram of van Boekel
\& Min (\citeyear{vboekelmin:2005}) for both the continuous
distribution of sizes (left) and the bimodal size distribution (right).
Models with the same size distribution are indicated by
lines. The different symbols are the different levels of turbulence.
\label{fig-result-boekel}}
\end{figure*}

We do the above described experiment for three values of the turbulent
strength: $\alpha=10^{-2}$, $10^{-3}$, $10^{-4}$, and for four different
continuous size distributions $p=-2.5$, $-2.0$, $-1.0$, $0.5$. The results
are shown in Fig.~\ref{fig-result-spec-distr}. We find that settling can
make the feature more pointy, but not dramatically so.  This is better seen
in Fig.~\ref{fig-result-boekel}-left, where we plot the well-known
strength-shape diagram of van Boekel \& Min
(\citeyear{vboekelmin:2005}). This diagram shows the strength
of the feature over the continuum on the x-axis and the normalized ratio of
the fluxes at 11.3 and 9.8 $\mu$m on the y-axis. Small grains (pointy features) appear at
the lower right of the diagram, while big grains (flat features) appear at the
top right. It should be noted here, however, that crystallization of the
grains have an effect similar to grain growth by ending up in the top left of
the diagram, but we are not concerned here about crystallization now. What
we do expect is that if settling causes a flat feature to be pointy, the
same model for ever lower values of $\alpha$ would be seen as a movement
from top left to bottom right. In the figure these models are connected
by lines. One sees that the expected effect is only reasonably strong for 
$p=-1$ and $p=0.5$, but weak or absent for $p=-2.5$ and $p=-2$. 

\begin{figure*}
\centerline{\includegraphics[width=0.8\textwidth]{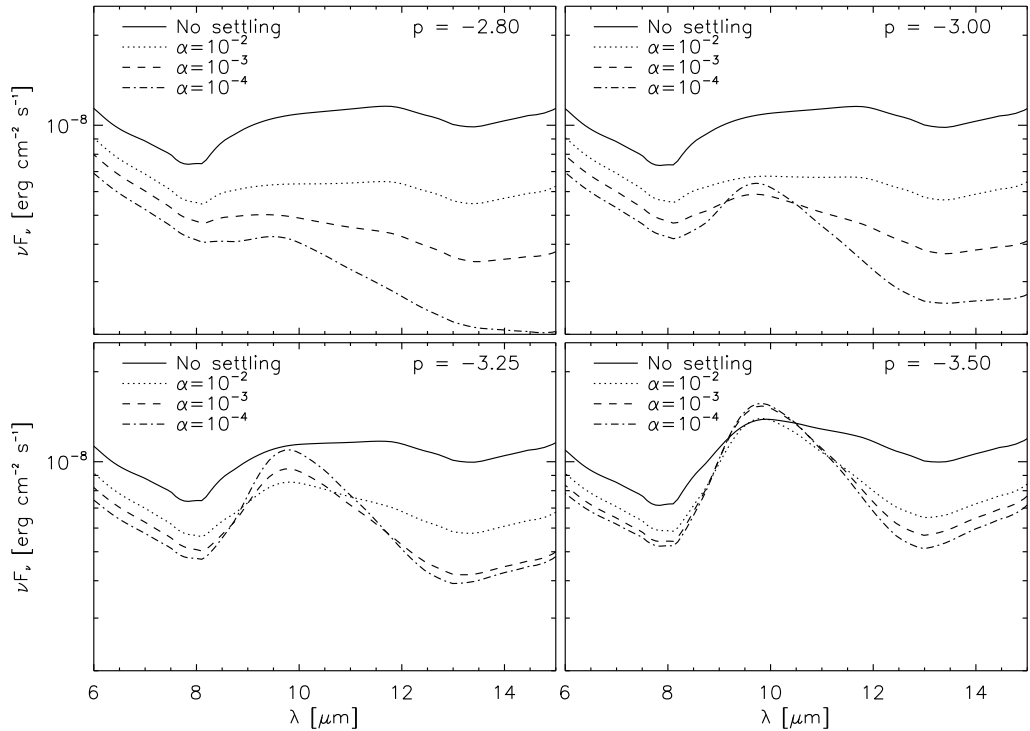}}
\caption{As Fig.~\ref{fig-result-spec-distr}. Model results for the bimodal
  size distribution with the small grain size 0.001 $\mu$m and the big grain
  size 3 $\mu$m.
\label{fig-result-spec-bimodal}}
\end{figure*}

Now we do the same for the bimodal size distribution, in which we have only
a small grain size (0.001 $\mu$m) and a big one ($3.0\mu$m).  Figure
\ref{fig-result-spec-bimodal} shows that the effect of settling is very
strong: a flat feature can now really be converted into a truly pointy
one. For $p=-2.8$, this happens quite suddenly between $\alpha=10^{-4}$ and
$\alpha=10^{-3}$, while for $p=-3.25$ this happens equally suddenly between
$\alpha=10^{-3}$ and $\alpha=10^{-2}$, and for $p=-3.5$ this happens for any
value of $\alpha$. This is also reflected in Fig.~\
\ref{fig-result-boekel}-right, where one can now clearly see the top-left
to bottom-right trend predicted when going from flat to
pointy. Interestingly, such a strong boosting of the small grain population
has also been seen in the models with PAHs (Dullemond et
al.~\citeyear{dullemond:2007}). So it appears that if we have a clearly
bimodal distribution, then the settling can boost the features of the small
grains (be they true grains or PAHs), but for a continuous distribution
this is not as easy.

\begin{figure*}
\includegraphics[width=\textwidth]{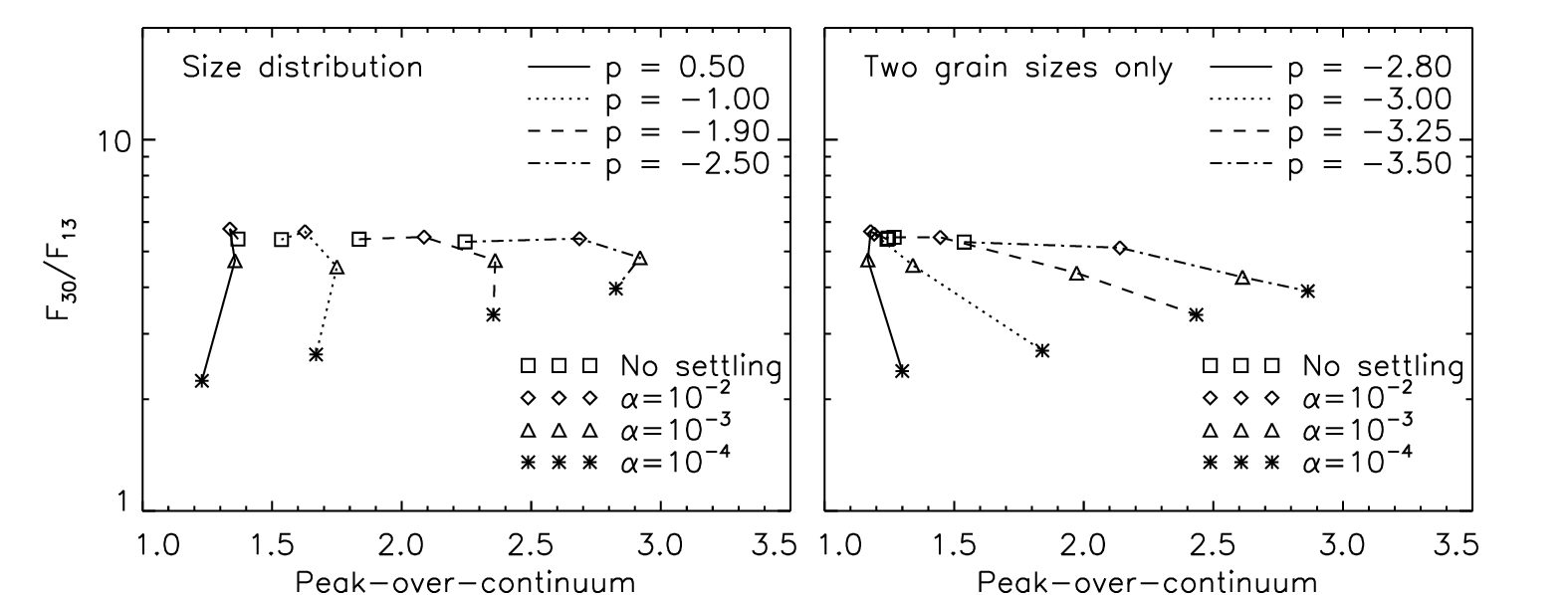}
\caption{The model predictions for the correlation between feature
strength (x-axis) and SED shape (y-axis). The SED shape is the 
ratio of the flux (in Jy) at 30 $\mu$m and at 13 $\mu$m. Left panel
is for the continuous size distribution models. Right panel is for
the bimodal size distribution models.
\label{fig-feature-sedshape}}
\end{figure*}

In addition to the feature-boosting effect described here, dust
sedimentation is also known to reduce the mid- to far-infrared flux of a
disk (Miyake \& Nakagawa \citeyear{miyakenaka:1995}; Chiang et
al.~\citeyear{chiangjoung:2001}; Dullemond \& Dominik
\citeyear{duldomsett:2004}; D'Alessio et
al.~\citeyear{dalessiocalvet:2006}). To be more precise: the ratio of, say,
30 $\mu$m flux over 13 $\mu$m flux will decrease for decreasing level of
turbulence (and hence increasing level of sedimentation). The reason for
this is that the disk becomes geometrically flatter as seen in the dust
continuum. The outer regions of the disk therefore capture less stellar
radiation and consequently produce weaker mid- to far-infrared
emission. Depending on the situation, it can even happen that the dust in
outer disk regions has sunk so deeply into the disk that it resides entirely
in the shadow cast by the inner disk regions (Dullemond \& Dominik
\citeyear{duldomsett:2004}). This suppresses the mid- to far-infrared flux
even more and also makes the disk nearly invisible in scattered light.  Now,
if sedimentation is responsible for the boosting of small-grain features in
some disks, then there should be a correlation between the feature
strength/shape and the mid- to far-infrared flux. In a recent paper, Bouwman
et al.~(\citeyear{bouwman:2008}) plot the flux ratio $F_{30\mu m}/F_{13\mu
  m}$ versus 10 $\mu$m feature-over-continuum for a small sample of T Tauri
stars. They find a correlation in which the disks with weak feature also
have a low $F_{30\mu m}/F_{13\mu m}$. They interpret this as the effect of
grain growth: as the grains grow, the 10 $\mu$m silicate feature becomes
weaker and flatter, and because the overall opacity of the disk goes down,
so does the mid- to far-infrared flux compared to the N-band flux. However,
in the scenario we study in this paper, in which different strengths of the
10 $\mu$m feature are supposed to be due to different strength of
turbulence, we would expect the opposite correlation: 
for low turbulence the feature becomes stronger, while the $F_{30\mu
  m}/F_{13\mu m}$ becomes weaker. Indeed this is what comes out of our
models, as can be seen in Fig.~\ref{fig-feature-sedshape}: A stronger
feature correlates with a smaller $F_{30\mu m}/F_{13\mu m}$. For the grain
size distribution this trend is rather weak (but certainly not the opposite
trend), but the trend is clear for the bimodal size distribution. If the
feature strength - SED shape correlation found by Boumwan et al.\ is
confirmed on the basis of larger samples, then this would serve as clear
proof that sedimentation alone cannot be held responsible for the strength
or weakness of the 10 $\mu$m feature.

\revised{In another recent paper, Watson et
  al.~(\citeyear{watsonleisen:2007}) found a correlation between the
  crystallinity of the dust as deduced from Spitzer IRS spectra and the SED
  overall shape. Disks that are more crystalline appear to have a flatter
  geometry. Since there appears to be no physical reason why crystalline
  grains are typically smaller than amorphous ones, the sedimentation
  filtering process we discussed in this paper does not predict any
  correlation between these things to happen. The explanation for the
  correlation found by Watson et al.\ must therefore lie in something
  else. In principle, as mentioned before, a perfect line-up of crystalline
  peaks in the 9-11 $\mu$m region of the spectrum may mimic a flat-topped
  feature from large amorphous silicates, and may affect the conclusions of
  Bouwman et al.~(\citeyear{bouwman:2008}). But it appears that the analysis
  by Bouwman et al.~is precise enough to exclude this possibility.}

\section{Conclusion}
\label{sec-conclusion}
We have modeled the sedimentation of grains of different sizes in a disk,
with particular emphasis on the 10 $\mu$m silicate feature shape.  We
expected that sedimentation removes big grains from the surface layers,
leaving the smaller grains behind, so that the spectrum will then be
dominated by a pointy 10 $\mu$m feature indicative of small grains, even if
the initial size distribution is dominated by the flat-topped features
characteristic of big (3$\mu$m) grains. We find that this is indeed the case
for {\em bimodal} size distributions if the smaller grains size is
chosen extremely small.  In our model  we have a clearly defined
population of $3\mu$m size grains and a clearly defined $\ll 0.1\mu$m
grains.  The arising effect is reasonably strong because there is a
large difference in the surface-to-mass ratio between these grains,
and therefore a strong effect on differential settling.

On the other hand, our attempt to produce this effect with a
continuous size distribution have basically failed.  With a continuous
(powerlaw) distribution of grains spanning the range $0.001\mu$m all
the way to $3\mu$m, we find only a very weak effect, much less than
what is observed. Therefore, unless the dust size distribution is strongly
bimodal distribution with just the right large and small grain sizes,
this excludes sedimentation alone as the sole origin of the different
10 $\mu$m silicate feature strengths and shapes seen in the spectra of
protoplanetary disks.  In other words, the shape and strength of the
silicate feature, which originates in the surface layers of the disk,
is reasonably representative of the dust hidden deep within the disk,
at least for grains in the size range up to a few micron.  Changes in
the 10\um feature do indicate changes in the small grain population in
the disk.

A similar conclusion can be made with respect to the influence of the
distribution of grains on the overall geometry of the disk.  A pure
sedimentation model (i.e.\ without growth) predicts that disks with low 30
$\mu$m flux over 13 $\mu$m flux ratio should typically have a stronger and
pointier 10 $\mu$m.  This is most clearly expected for the \emph{bimodal}
size distribution.  For the \emph{continuous} size distribution, this
correlation is expected to be weaker because of the weaker effect of
sedimentation on the silicate feature shape.  But at the very least such a
model does {\em not} predict the \emph{opposite} trend: that weak 30-over-13
flux ratio correlates with a weak feature, which is the trend observed by
Bouwman et al.~(\citeyear{bouwman:2008}). If this trend can be confirmed for
a larger sample, this would then constitute rather strong proof that
sedimentation alone cannot be held responsible for the variety in observed
silicate strengths.  Dust coagulation, in combination of course with
sedimentation, must then play a vital role in producing the different
silicate strengths and shapes observed in disks around young stars.

\begin{acknowledgements}
  We wish to thank Rens Waters, Thomas Henning, and Jeroen Bouwman for very
  stimulating discussions.
\end{acknowledgements}


\end{document}